\documentclass[letterpaper]{article} % DO NOT CHANGE THIS
\usepackage{aaai24}  % DO NOT CHANGE THIS
\usepackage{times}  % DO NOT CHANGE THIS
\usepackage{helvet}  % DO NOT CHANGE THIS
\usepackage{courier}  % DO NOT CHANGE THIS
\usepackage[hyphens]{url}  % DO NOT CHANGE THIS
\usepackage{graphicx} % DO NOT CHANGE THIS
\usepackage{natbib}  % DO NOT CHANGE THIS AND DO NOT ADD ANY OPTIONS TO IT
\usepackage{caption} % DO NOT CHANGE THIS AND DO NOT ADD ANY OPTIONS TO IT
\usepackage{soul}
\usepackage[utf8]{inputenc}
\usepackage{amsmath}
\usepackage{amsthm}
\usepackage{booktabs}
\usepackage[switch]{lineno}
\usepackage{amssymb,subcaption,multirow,dsfont,epsfig,bm }
\usepackage{makecell}

\newtheorem{theo}{Theorem}
\newtheorem{lemm}{\textbf{lemma}}

\usepackage{algorithm}
\usepackage{algorithmic}

\usepackage{newfloat}
\usepackage{listings}
\DeclareCaptionStyle{ruled}{labelfont=normalfont,labelsep=colon,strut=off} % DO NOT CHANGE THIS
\floatstyle{ruled}
\newfloat{listing}{tb}{lst}{}
\floatname{listing}{Listing}
\title{Learning Diffusions under Uncertainty}

\author {
    Hao Huang\textsuperscript{\rm 1},
    Qian Yan\textsuperscript{\rm 1},
    Keqi Han\textsuperscript{\rm 2},
    Ting Gan\textsuperscript{\rm 1},
    Jiawei Jiang\textsuperscript{\rm 1},
    Quanqing Xu\textsuperscript{\rm 3},
    Chuanhui Yang\textsuperscript{\rm 3}
}
\affiliations {
    \textsuperscript{\rm 1}School of Computer Science, Wuhan University, China\\
    \textsuperscript{\rm 2}Department of Computer Science, Emory University, USA\\
    \textsuperscript{\rm 3}OceanBase, AntGroup, China\\
    \{haohuang, qy, ganting, jiawei.jiang\}@whu.edu.cn, keqi.han@emory.edu, \{xuquanqing.xqq, rizhao.ych\}@oceanbase.com
}

\begin{document}

\maketitle

\begin{abstract}
 To infer a  diffusion network based on observations from historical diffusion processes,
  existing approaches assume that observation data contain exact occurrence time of each node infection,
  or at least the eventual infection statuses of nodes in each diffusion process. They determine potential influence relationships between nodes by
  identifying frequent sequences,  or statistical correlations, among node infections.
  In some real-world settings, such as the spread of epidemics, tracing exact infection times is often infeasible due to a high cost; even obtaining precise infection statuses of nodes is a challenging task, since
   observable symptoms such as headache only partially reveal a node's true status.
  In this work, we investigate  how to effectively infer a diffusion network from observation data with uncertainty.
  Provided with only probabilistic information about
  node infection statuses, we formulate the problem of diffusion network inference as a constrained nonlinear regression w.r.t.~the probabilistic data.
  An alternating maximization method is designed to  solve this regression problem iteratively, and the improvement of solution  quality in each iteration can be theoretically guaranteed.
  Empirical studies are conducted on both synthetic and real-world networks, and the results  verify the effectiveness and efficiency of our approach.
  
\end{abstract}

\section{Introduction}

The spread of viewpoints, rumors, and diseases are often modelled as probabilistic processes over a diffusion network.
In the network, a directed edge represents a parent-child influence relationship, which
 indicates that the parent node can influence the child node with a certain probability.
%%%%
In most cases, such influence relationships are not naturally visible or traceable, and
we only observe a set of  diffusion results from a limited number of historical  diffusion  processes \cite{poind}.
Diffusion network inference aims to infer the diffusion network structure (i.e., the topology of influence relationships)
from the observation data.
This problem  has received considerable attention  
in areas 
such as information propagation \cite{he2},
viral marketing \cite{copurch},  and epidemic prevention \cite{epidemic}
% such as social networks,   viral marketing  and
% epidemic prevention, % \cite{grouplevel}, 
since the inferred diffusion network structure enables an
intuitive understanding of the underlying interactions between
nodes, and is essential for developing strategies to control future diffusion processes \cite{sidn}.

Most existing approaches  %to diffusion network inference 
assume that
observation data contain  the exact times when node infections occurred.
With this temporal information, they
determine  potential influence relationships between nodes
by identifying frequent sequences of node infections,
since nodes infected sequentially within a short time interval are considered more likely to possess influence relationships \cite{1,2,3,4,5,7,8,9,11}.
Nonetheless,
 monitoring every  node constantly during each diffusion process
to obtain such temporal information of node infections is often expensive    in practice.
Therefore,
some other approaches aim to
carry out diffusion network inference without temporal information, using only the eventual
infection statuses of nodes observed at the end of each diffusion process \cite{06,twind,sidn,ijcai22,lmdn,tends,poind}.
Toward this, they measure the statistical correlations  of node infections, and
identify influence relationships by checking which node pairs have high
infection correlations.

In this paper, we study the problem of diffusion network inference
in a less idealized and more realistic setting, i.e.,
% we are provided only with probabilistic information about
%  node infection statuses.
only probabilistic information about node infection statuses is provided.
This uncertainty  is
interpreted differently in various contexts.
For example, in epidemic containment,
it is difficult to  confirm  the
infection statuses of outpatients  based on observable symptoms such as headache
and fatigue, since there is a certain probability that
these symptoms may be caused by   other reasons like lack of sleep;
in viral marketing campaigns, it is often the case that
the  respondents are more prone to  probabilistic  feedback
than a clear-cut answer.
To the best of our knowledge,
 only one existing work  \cite{12} has partially
addressed the problem of inferring
diffusion networks based on probabilistic data.
In addition to the probabilistic information about
 node infection statuses,
 that work further requires   prior knowledge on the transmission probabilities between
 different node  statuses, and the   temporal information of node infections.

Aiming at a more general solution to inferring diffusion networks from probabilistic data,
we formulate the problem as a constrained nonlinear regression w.r.t.~the probabilistic data,
and propose an effective and efficient algorithm called
  {PIND} (a re-ordered acronym of \textbf{I}nferring \textbf{N}etworks from  \textbf{P}robabilistic \textbf{D}ata)
to solve the regression problem iteratively.
In each iteration, %  of the algorithm,
an alternating maximization method is adopted to
 update the estimate on the probability of the existence of each potential influence relationship,
 and  the  influence relationship strengthens.
 The quality of the estimate is theoretically guaranteed to improve  across the iterations.
After the convergence of iterations,
the most probable influence relationships in the objective diffusion network can be easily identified based on the estimate. % the probability estimation result.

The remainder of the paper is organized as follows. We
first present our problem statement, and then introduce our
proposed PIND algorithm, followed by reporting experimental
results and our findings before concluding the paper.

\section{Problem Statement}

A diffusion network is represented as a
directed graph $G=(V,E)$, where   $V=\{v_1,..., v_n\}$ is the set of $n$ nodes in the network, and  $E$ is the set of
directed edges (i.e., influence relationship) between the nodes.
A directed edge $(v_j, v_i)\in E$ from a parent node $v_j\in V$ to a child node $v_i\in V$ indicates that
when $v_j$ is infected and $v_i$ is uninfected,
$v_j$ will infect $v_i$ with a certain probability $\alpha_{ji}$, which is known as  infection propagation  probability and can be
regarded as the strength of the influence relationship from $v_j$ to $v_i$.
We assume that the diffusion processes on $G$ follow the Independent Cascade (IC) model,
in which each infected parent node tries to infect each of its uninfected children
with corresponding infection propagation  probability independently.
This assumption is customarily adopted in prior art, and we would like to point out that
our proposed algorithm can be extended to other propagation models such as the linear threshold model.

In the problem of  diffusion network inference,
the node set $V$ is given, while the network structure, i.e., directed edge set $E$, %($t \in \{1,...,\alpha\}$),
is unknown, neither the infection propagation probabilities w.r.t.~the structure.
To infer the structure of a diffusion network, a set $S$ of diffusion results observed from a number of historical diffusion processes
on the network is required.
%%%
In this paper, we assume that the diffusion results $S$ contain only the
probabilistic information
of node infection statuses, i.e., $S=\{s^\ell_{i}   \mid i\in\{1,...,n\}, \ell\in\{1,...,\beta\}\}$, where
 $\beta$ is the number of historical diffusion processes, and
$s_i^\ell \in [0,1]$ refers to the
probability that node $v_i$ is infected  in the $\ell$-th  historical diffusion process.
%%%
Then, our  problem statement for diffusion network inference with probabilistic data can be formulated as follows.

  \textbf{\emph{Given}:} a set   $S = \{ {S^1},...,{S^\beta}\}$ of  diffusion results observed
on a  diffusion network $G$ at the end of  $\beta$   diffusion processes,
where  ${S^\ell } = \{s_1^\ell ,...,s_n^\ell \}$ ($\ell \in \{1,...,\beta\}$)
 records the   infection probability  $s_i^\ell $   of each node $v_i\in V$   in the $\ell$-th diffusion process.

  \textbf{\emph{Infer}:}  the edge set $E$ of the diffusion network $G$.

\section{The PIND Algorithm}

In this section, we first elaborate how to
estimate the existence of   influence relationships  in a probabilistic way,
and then present how to reduce redundant computation
in the influence relationship estimation with a pruning method for candidate parent nodes. We
conclude this section with a complexity analysis on our approach.

\subsection{Estimation of Influence Relationships}

\subsubsection{Objective  Function}

Let $F_i$ be the set of parent nodes of node $v_i\in V$ in objective diffusion network,
according to IC model,
the log-likelihood of $v_i$ being infected by any node in set $F_i$ in
the $\ell$-th diffusion process can be calculated as
\begin{equation}
   \log(1-\prod_{v_j\in F_i} (1-s_j^\ell \alpha_{ji})),
\end{equation}
the log-likelihood of $v_i$ not being infected by any node in set $F_i$ in
the $\ell$-th diffusion process can be calculated as
\begin{equation}
   \log(\prod_{v_j\in F_i} (1-s_j^\ell \alpha_{ji}))=\sum_{v_j\in F_i}\log(1-s_j^\ell \alpha_{ji}).
\end{equation}
Since the probability that node $v_i$ is infected in the $\ell$-th historical diffusion process is $s_i^\ell$ and
the probability that node $v_i$ is not infected in the $\ell$-th historical diffusion process is $1-s_i^\ell$,
then the log-likelihood of observation $s_i^\ell$ is
\begin{equation} \label{eq:like}
\begin{aligned}
  \mathcal{G}(s_i^\ell)
  =&s_i^\ell \log\big(1-\prod_{v_j\in F_i} (1-s_j^\ell \alpha_{ji})\big) +\\
  &(1-s_i^\ell)\sum_{v_j\in F_i}\log(1-s_j^\ell \alpha_{ji}).
\end{aligned}
\end{equation}
%%%
%%%%%%
Let variable $x_{ji}\in\{0,1\}$ indicate whether there is a directed edge from node $v_j$ to node $v_i$ (1 for yes and 0 for no),
and $C_i = V\setminus \{v_i\}$ denote the set of all possible candidate parent nodes of $v_i$
if we can infer each variable $x_{ji}$ ($v_i\in V, v_j \in C_i$) accurately,
then according to Eq.~(\ref{eq:like}), the following equation should be satisfied.
%%%%%%
\begin{equation} \label{eq:like:sil}
\begin{aligned}
  \mathcal{G}(s_i^\ell)
  =&s_i^\ell \log\big(1-\prod_{v_j\in C_i} (1-s_j^\ell \alpha_{ji})^{x_{ji}}\big) +\\
  &(1-s_i^\ell)\sum_{v_j\in C_i}{x_{ji}}\log(1-s_j^\ell \alpha_{ji}).
\end{aligned}
\end{equation}
%%%
%%%
Let $x$ be the collection of all $x_{ji}$ ($v_i\in V, v_j \in C_i$), 
and $\alpha$ be the collection of all $\alpha_{ji}$ ($v_i\in V, v_j \in C_i$),
then the overall log-likelihood of data $S$ is
% \begin{equation}
% \begin{aligned}
%   \mathcal{G}(S) =
%   &\sum_{\ell=1}^\beta \sum_{i=1}^n s_i^\ell \log\big(1-\prod_{v_j\in C_i} (1-s_j^\ell \alpha_{ji})^{x_{ji}}\big) +\\
%   &\sum_{\ell=1}^\beta \sum_{i=1}^n (1-s_i^\ell)\sum_{v_j\in C_i}{x_{ji}}\log(1-s_j^\ell \alpha_{ji}),
% \end{aligned}
% \end{equation}
%
\begin{equation}
\begin{aligned}
  \mathcal{G}(S) =
  &\sum_{\ell=1}^\beta \sum_{i=1}^n \mathcal{G}(s_i^\ell), %s_i^\ell \log\big(1-\prod_{v_j\in C_i} (1-s_j^\ell \alpha_{ji})^{x_{ji}}\big) +\\
  %&\sum_{\ell=1}^\beta \sum_{i=1}^n (1-s_i^\ell)\sum_{v_j\in C_i}{x_{ji}}\log(1-s_j^\ell \alpha_{ji}),
\end{aligned}
\end{equation}
%%%
%%%%%%%
which is a function w.r.t. $x$ and $\alpha$.
Let
\begin{equation}
  \mathcal{L}(x,\alpha)=\mathcal{G}(S).
\end{equation}
%%%%%
A greater value of   $\mathcal{L}(x,\alpha)$ indicates that the corresponding  $x$ and $\alpha$ are
more close to the truth. Therefore, our goal is to find optimal  $x$ and $\alpha$ that
 maximize the value of $\mathcal{L}(x,\alpha)$, which can be formulated as follows.
%\begin{equation} \label{problem1}
%\begin{aligned}
%  \max\mathcal{L}=
%  &\sum_{\ell=1}^\beta \sum_{i=1}^n s_i^\ell \big(1-\prod_{v_j\in C_i} (1-s_j^\ell \alpha_{ji})^{x_{ji}}\big) +\\
%  &\sum_{\ell=1}^\beta \sum_{i=1}^n (1-s_i^\ell)\prod_{v_j\in C_i} (1-s_j^\ell \alpha_{ji})^{x_{ji}}, \\
%  \mbox{s.t.}&~  x_{ji}\in\{0,1\}, \alpha_{ji}\in[0,1], \forall i,j.
%\end{aligned}
%\end{equation}
\begin{equation} \label{problem1}
\begin{aligned}
  \max\mathcal{L}=
  &\sum_{\ell=1}^\beta \sum_{i=1}^n s_i^\ell \log\big(1-\prod_{v_j\in C_i} (1-s_j^\ell \alpha_{ji})^{x_{ji}}\big) +\\
  &\sum_{\ell=1}^\beta \sum_{i=1}^n (1-s_i^\ell)\sum_{v_j\in C_i}{x_{ji}}\log(1-s_j^\ell \alpha_{ji}) \\
  \mbox{s.t.}&~  x_{ji}\in\{0,1\}, \alpha_{ji}\in[0,1], \forall i,j.
\end{aligned}
\end{equation}
The constrained maximization problem above is a nonlinear mixed integer programming, which is difficult to be solved directly.
Therefore, we relax the integer constraint on each $x_{ji}$ by allowing $x_{ji}\in[0,1]$ ($j,i\in\{1,...,n\}$)
such that continuous programming methods
can be applied to this problem, where the continuous value of
each $x_{ji}$ denotes the probability that there is a directed edge from node $v_j$ to node $v_i$.
This kind of relaxations are commonly used to solve the problem of integer programming.
Then, the  problem in Eq.~(\ref{problem1}) is relaxed and reformulated as follows.
%Then, the problem in Eq.~(\ref{problem1}) is relaxed as follows.
%\begin{equation} \label{opt:gold}
%\begin{aligned}
%  \max\mathcal{L}=
%  &\sum_{\ell=1}^\beta \sum_{i=1}^n s_i^\ell \big(1-\prod_{v_j\in C_i} (1-s_j^\ell \alpha_{ji})^{x_{ji}}\big) +\\
%  &\sum_{\ell=1}^\beta \sum_{i=1}^n (1-s_i^\ell)\prod_{v_j\in C_i} (1-s_j^\ell \alpha_{ji})^{x_{ji}}, \\
%  \mbox{s.t.}&~  x_{ji}\in[0,1], \alpha_{ji}\in[0,1], \forall i,j.
%\end{aligned}
%\end{equation}
\begin{equation} \label{opt:gold}
\begin{aligned}
  \max\mathcal{L}=
  &\sum_{\ell=1}^\beta \sum_{i=1}^n s_i^\ell \log\big(1-\prod_{v_j\in C_i} (1-s_j^\ell \alpha_{ji})^{x_{ji}}\big) +\\
  &\sum_{\ell=1}^\beta \sum_{i=1}^n (1-s_i^\ell)\sum_{v_j\in C_i}{x_{ji}}\log(1-s_j^\ell \alpha_{ji}) \\
  \mbox{s.t.}&~  x_{ji}\in[0,1], \alpha_{ji}\in[0,1], \forall i,j.
\end{aligned}
\end{equation}

\subsubsection{Solving Method}

To solve the problem in Eq.~(\ref{opt:gold}),
the straightforward method is solving equations $\frac{\partial \mathcal{L} }{\partial x}=0$ and $\frac{\partial \mathcal{L} }{\partial \alpha  }=0$, where $\frac{\partial \mathcal{L} }{\partial x}$ and $\frac{\partial \mathcal{L} }{\partial \alpha  }$ are the derivatives of
$\mathcal{L}(x,\alpha)$ w.r.t.~$x$ and $\alpha$, respectively, which can be calculated as
\begin{align} \label{eq:pij}
  \frac{\partial \mathcal{L} }{\partial \alpha_{ji} }= \sum_{\ell=1}^\beta \frac{x_{ji}s_j^\ell}{1-s_j^\ell \alpha_{ji}} L_i^\ell,
\end{align}
%%%%
%%%%
\begin{align} \label{eq:xij}
  \frac{\partial \mathcal{L} }{\partial x_{ji} }
  =   -\sum_{\ell=1}^\beta \log (1-s_j^\ell \alpha_{ji}) L_i^\ell.
\end{align}
where
\begin{align} \label{eq:xij2}
L_i^\ell = s_i^\ell\frac{\prod_{v_k\in C_i} (1-s_k^\ell \alpha_{ki})^{x_{ki}}}{1-\prod_{v_k\in C_i} (1-s_k^\ell \alpha_{ki})^{x_{ki}} }-(1-s_i^\ell)  .
\end{align}

Nevertheless, directly solving equations $\frac{\partial \mathcal{L} }{\partial x}=0$ is still difficult, and its solution
may not satisfy the  constraints in Eq.~(\ref{opt:gold}).
To address this issue, we adopt an alternating maximization method, which works as follows.

\textbf{Step 1.} Initializing $\alpha$ and $x$ under   constraints that  $x_{ji}\in [0,1]$, $\alpha_{ji} \in [0,1]$, $\forall i,j$.

\textbf{Step 2.} %Submitting   current $\alpha$  into Eq.~(\ref{eq:xij}) and solving $\frac{\partial \mathcal{L} }{\partial x }=0$ for $x$. %Since $\frac{\partial \mathcal{L} }{\partial x }$ is linear w.r.t.~$x$, it can be solved directly. Let $x'$ be the solution,  if there doesn't exist $x'_{ji}$ such that $x'_{ji}<0$ or $x'_{ji}>1$ then update the current $x$'s as the solution $x'$.
Selecting an appropriate update direction $y$ (based on $\frac{\partial \mathcal{L} }{\partial x }$) and an appropriate step size $\theta$, and updating $x$ as
\begin{align} \label{eq:step2}
  x \gets x+\theta y.
\end{align}

\textbf{Step 3.} Selecting an appropriate update direction $z$ (based on $\frac{\partial \mathcal{L} }{\partial \alpha }$) and an appropriate step size $\lambda$, and  updating $\alpha$ as
\begin{align} \label{eq:step3}
   \alpha \gets \alpha+\lambda z.
\end{align}

\textbf{Step 4.} Repeating Steps 2 and 3 until convergence.

In the above alternating maximization method, how to
selecting appropriate update directions and  step sizes for $x$ and $\alpha$  is of central importance.
Let $x^{(T)}$ be the value of $x$, and $\alpha^{(T)}$ be the value of $\alpha$, in the $T$-th iteration of Steps 2 and 3,
$f(x)=\mathcal{L}(x, \alpha^{(T)})$ be a function w.r.t.~$x$,
and  $g(x) = \mathcal{L}(x^{(T)}, \alpha)$ be a function w.r.t.~$\alpha$,
then we have
\begin{equation}
  \frac{\partial f }{\partial x } =\frac{\partial \mathcal{L} }{\partial x }\mid_{\alpha=\alpha^{(T)}},
\end{equation}
%%%
\begin{equation}
  \frac{\partial g }{\partial \alpha } =\frac{\partial \mathcal{L} }{\partial \alpha }\mid_{x=x^{(T)}}.
\end{equation}

As the direction of gradient is the steepest ascent direction,     we can utilize the gradient to  update $x$ and $\alpha$.  %direction $y$ and $z$ based on
 Furthermore, to make the    constraints in Eq.~(\ref{opt:gold})  satisfied,
we modify  the  gradient and select the update directions $y$ and $z$ of variables $x$ and $\alpha$ as follows.
\begin{equation}\label{eq:yji}
y_{ji}\!=\!\left\{
  \begin{aligned}
   & 0,~\mbox{ if } x_{ji}^{(T)}\!=\!0, \frac{\partial \mathcal{L} }{\partial x_{ji} }\mid_{(x,\alpha)=(x^{(T)},\alpha^{(T)})}<0;\\
   & 0,~\mbox{ if } x_{ji}^{(T)}\!=\!1, \frac{\partial \mathcal{L} }{\partial x_{ji} }\mid_{(x,\alpha)=(x^{(T)},\alpha^{(T)})}>0;\\
   & \frac{\partial \mathcal{L} }{\partial x_{ji} }\mid_{(x,\alpha)=(x^{(T)},\alpha^{(T)})},~~otherwise.
  \end{aligned}
\right.
\end{equation}
\begin{equation}\label{eq:zji}
z_{ji}\!=\!\left\{
  \begin{aligned}
   & 0,~\mbox{if } \alpha_{ji}^{(T)}\!=\!0, \frac{\partial \mathcal{L} }{\partial \alpha_{ji} }\mid_{(x,\alpha)=(x^{(T)},\alpha^{(T)})}<0;\\
   & 0,~\mbox{if } \alpha_{ji}^{(T)}\!=\!1, \frac{\partial \mathcal{L} }{\partial \alpha_{ji} }\mid_{(x,\alpha)=(x^{(T)},\alpha^{(T)})}>0;\\
   & \frac{\partial \mathcal{L} }{\partial \alpha_{ji} }\mid_{(x,\alpha)=(x^{(T)},\alpha^{(T)})},~~otherwise.
  \end{aligned}
\right.
\end{equation}

Next, we discuss how to select step sizes $\theta$ and $\lambda$.
To this end, first, we define  $\theta_{ji}$ and $\lambda_{ji}$ as follows.
\begin{equation} \label{eq:theta}
 \theta_{ji}=\left\{
  \begin{aligned}
    \frac{x_{ji}}{-y_{ji}}, &&\mbox{ if } y_{ji}<0;\\
    \frac{1-x_{ji}}{y_{ji}}, &&\mbox{ if } y_{ji}>0;\\
    +\infty,&&\mbox{ if } y_{ji}=0.
  \end{aligned}
\right.
\end{equation}
\begin{equation} \label{eq:lembda}
 \lambda_{ji}=\left\{
  \begin{aligned}
    \frac{\alpha_{ji}}{-z_{ji}}, &&\mbox{ if } z_{ji}<0;\\
    \frac{1-\alpha_{ji}}{z_{ji}}, &&\mbox{ if } z_{ji}>0;\\
    +\infty,&&\mbox{ if } z_{ji}=0.
  \end{aligned}
\right.
\end{equation}
%%%%
%%%%
Let
\begin{equation}
  \theta'=\min\{\theta_{ji}\mid i\in \{1,...,n\}, v_j\in C_i\},
 \end{equation}
  relationship
$\theta \leqslant \theta'$   can guarantee that if $x^{(T)}$ is feasible (i.e., $x_{ji}^{(T)}\in [0,1]$, $\forall i,j$),
then  $x^{(T)}+\theta y$ is also feasible.
Similarly, let
\begin{equation}
 \lambda'=\min\{\lambda_{ji}\mid i\in \{1,...,n\}, v_j\in C_i\},
 \end{equation}
 relationship $\lambda \leqslant \lambda'$  can guarantee that if
  $\alpha^{(T)}$ is feasible,  then $\alpha^{(T)}+\lambda z$ is also feasible.
In other words, $\theta'$ and $\lambda'$ should be the upper bounds of step sizes $\theta$ and $\lambda$, respectively.
Furthermore, to guarantee
the value of objective function $\mathcal{L}$
becoming  greater after the updating of $x$ and $\alpha$, we
could select step sizes $\theta$ and $\lambda$  based on the following two lemmas.
%Note that all the proofs related to this section are given in the supplementary material \cite{}.

\begin{lemm}\label{lemma:1}
    If $y\neq 0$  (i.e., $\exists y_{ji}\neq 0$), then
  there exists a  nonnegative integer $m$ such that $f(x^{(T)}+\frac{\theta'}{2^m} y)> f(x^{(T)})$.
\end{lemm}
 \textbf{Proof:}
Firstly, we prove that when $y\neq 0$,  relationship  $0< \theta' < +\infty$ holds.

If $y_{ji}<0$, i.e.,
%\begin{equation}
 $ \frac{\partial \mathcal{L} }{\partial x_{ji} }\mid_{(x,\alpha)=(x^{(T)},\alpha^{(T)})} <0$,
%\end{equation}
then $x^{(T)}_{ji}\neq 0$ should be satisfied, otherwise
%%%
according to definition of $y_{ji}$,    the value of $y_{ji}$ should be 0, which contradicts with the condition $y_{ji}<0$.
As the feasible region of $x^{(T)}_{ji}$ is $[0,1]$, when $y_{ji}<0$,
  $x^{(T)}_{ji} > 0$ holds, so $\theta_{ji} = \frac{x^{(T)}_{ji}}{-y_{ji}} >0$.

If $y_{ji}>0$, i.e.,
$ \frac{\partial \mathcal{L} }{\partial x_{ji} }\mid_{(x,\alpha)=(x^{(T)},\alpha^{(T)})} >0$,
then $x^{(T)}_{ji}\neq 1$ should be satisfied, otherwise
%%%
according to definition of $y_{ji}$, the value of $y_{ji}$ should be 0, which contradicts with the condition $y_{ji}>0$.
As the feasible region of $x^{(T)}_{ji}$ is $[0,1]$, when $y_{ji}>0$,
  $  x^{(T)}_{ji}< 1$ holds, so $\theta_{ji} = \frac{1- x^{(T)}_{ji}}{y_{ji}} >0$.

If $y_{ji} = 0$, according to definition of $\theta_{ji}$, $\theta_{ji} > 0$ holds.

In brief, $\theta_{ji} > 0$ always holds.
Moreover, since $\exists y_{ji}\neq 0$, according to the definitions of $\theta_{ji}$ and $\theta'$,  we have
$
  \theta' \leqslant \theta_{ji} < +\infty
$. Thus, when $y\neq 0$,  relationship  $0< \theta' < +\infty$ holds.

Secondly, we prove that if  $y\neq 0$  and  $0< \theta' < +\infty$, then
there exists an nonnegative integer $m$ such that $f(x^{(T)}+\frac{\theta'}{2^m} y)> f(x^{(T)})$.

Let
%\begin{equation}
 $h(\theta)=f(x^{(T)}+\theta y)$, then
%\end{equation}
its derivative is
\begin{equation}
\begin{aligned}
  h'(\theta)=& y\cdot \frac{\partial f }{\partial x }\mid_{x^{(T)}+\theta y}  \\
  =&y\cdot\frac{\partial \mathcal{L} }{\partial x }\mid_{(x,\alpha)=(x^{(T)}+\theta y,\alpha^{(T)})}.
\end{aligned}
\end{equation}
Since $y\neq 0$, we have
\begin{equation}
%\begin{aligned}
  h'(0)= y\cdot\frac{\partial \mathcal{L} }{\partial x }\mid_{(x,\alpha)=(x^{(T)},\alpha^{(T)})}
  =y\cdot y >0.
%\end{aligned}
\end{equation}
As $h'(\theta)$ is continuous, there exists $\delta>0$ such that
\begin{equation} \label{eq22}
  \forall 0\leqslant\theta\leqslant \delta,~~h'(\theta)>0.
\end{equation}
As $0< \theta' < +\infty$,
  there exists a great enough integer $m$   such that
\begin{equation} \label{eq24}
  0< \frac{\theta'}{2^m}\leqslant\delta.
\end{equation}
Combining Eqs.~(\ref{eq22})~\&~(\ref{eq24}), we have
\begin{equation} \label{eq25}
  \forall 0\leqslant\theta\leqslant \frac{\theta'}{2^m}, h'(\theta)>0.
\end{equation}
According Eq.~(\ref{eq25}),
 we have  $h(\frac{\theta'}{2^m})>h(0)$, which is equivalent to
  $f(x^{(T)}+\frac{\theta'}{2^m} y)> f(x^{(T)})$.

 Therefore,  there exists a  nonnegative integer $m$ such that $f(x^{(T)}+\frac{\theta'}{2^m} y)> f(x^{(T)})$,
and Lemma 1 is correct.
\hfill $\blacksquare$

\begin{lemm}\label{lemma:2}
  If $z\neq 0$ (i.e., $\exists z_{ji}\neq 0$), then
  there exist a  nonnegative integer $k$ such that $g(\alpha^{(T)}+\frac{\lambda'}{2^k} z)> g(\alpha^{(T)})$.
\end{lemm}
  \textbf{Proof:} 
Firstly, we prove that when $z\neq 0$,  relationship  $0< \lambda' < +\infty$ holds.

If $z_{ji}<0$, i.e.,
%\begin{equation}
 $ \frac{\partial \mathcal{L} }{\partial \alpha_{ji} }\mid_{(x,\alpha)=(x^{(T)},\alpha^{(T)})} <0$,
%\end{equation}
then $\alpha^{(T)}_{ji}\neq 0$ should be satisfied, otherwise
%%%
according to definition of $z_{ji}$,    the value of $z_{ji}$ should be 0, which contradicts with the condition $z_{ji}<0$.
As the feasible region of $\alpha^{(T)}_{ji}$ is $[0,1]$, when $z_{ji}<0$,
  $\alpha^{(T)}_{ji} > 0$ holds, so $\lambda_{ji} = \frac{\alpha^{(T)}_{ji}}{-z_{ji}} >0$.

If $z_{ji}>0$, i.e.,
$ \frac{\partial \mathcal{L} }{\partial \alpha_{ji} }\mid_{(x,\alpha)=(x^{(T)},\alpha^{(T)})} >0$,
then $\alpha^{(T)}_{ji}\neq 1$ should be satisfied, otherwise
%%%
according to definition of $z_{ji}$, the value of $z_{ji}$ should be 0, which contradicts with the condition $z_{ji}>0$.
As the feasible region of $\alpha^{(T)}_{ji}$ is $[0,1]$, when $z_{ji}>0$,
  $  \alpha^{(T)}_{ji}< 1$ holds, so $\lambda_{ji} = \frac{1- \alpha^{(T)}_{ji}}{z_{ji}} >0$.

If $z_{ji} = 0$, according to definition of $\lambda_{ji}$, $\lambda_{ji} > 0$ holds.

In brief, $\lambda_{ji} > 0$ always holds.
Moreover, since $\exists z_{ji}\neq 0$, according to the definitions of $\lambda_{ji}$ and $\lambda'$,  we have
$
  \lambda' \leqslant \lambda_{ji} < +\infty
$. Thus, when $z\neq 0$,  relationship  $0< \lambda' < +\infty$ holds.

Secondly, we prove that if  $z \neq 0$  and  $0< \lambda' < +\infty$, then
there exists an nonnegative integer $k$ such that $g(\alpha^{(T)}+\frac{\lambda'}{2^k} z)> g(\alpha^{(T)})$.

Let
%\begin{equation}
 $H(\lambda)=g(\alpha^{(T)}+\lambda z)$, then
%\end{equation}
its derivative is
\begin{equation}
\begin{aligned}
  H'(\lambda)=& z\cdot \frac{\partial g }{\partial \alpha }\mid_{\alpha^{(T)}+\lambda z}  \\
  =&z\cdot\frac{\partial \mathcal{L} }{\partial \alpha }\mid_{(x,\alpha)=(x^{(T)},\alpha^{(T)}+\lambda z)}.
\end{aligned}
\end{equation}
Since $z\neq 0$, we have
\begin{equation}
%\begin{aligned}
  H'(0)= z\cdot\frac{\partial \mathcal{L} }{\partial \alpha }\mid_{(x,\alpha)=(x^{(T)},\alpha^{(T)})}
  =z\cdot z >0.
%\end{aligned}
\end{equation}
As $H'(\lambda)$ is continuous, there exists $\sigma>0$ such that
\begin{equation} \label{eq32}
  \forall 0\leqslant\lambda\leqslant \sigma,~~H'(\lambda)>0.
\end{equation}
As $0< \lambda' < +\infty$,
  there exists a great enough integer $k$   such that
\begin{equation} \label{eq34}
  0< \frac{\lambda'}{2^k}\leqslant\sigma.
\end{equation}
Combining Eqs.~(\ref{eq32})~\&~(\ref{eq34}), we have
\begin{equation} \label{eq35}
  \forall 0\leqslant\lambda\leqslant \frac{\lambda'}{2^k}, H'(\lambda)>0.
\end{equation}
According Eq.~(\ref{eq35}),
 we have  $H(\frac{\lambda'}{2^k})>H(0)$, which is equivalent to
  $g(\alpha^{(T)}+\frac{\lambda'}{2^k} z)> g(\alpha^{(T)})$.

 Therefore,  there exists a  nonnegative integer $k$ such that $g(\alpha^{(T)}+\frac{\lambda'}{2^k} z)> g(\alpha^{(T)})$,
and Lemma 2 is correct.  \hfill $\blacksquare$

% The proof process  is the same as the proof process of Lemma \ref{lemma:1}.
% \hfill $\blacksquare$

Based on Lemmas~ \ref{lemma:1}~\&~\ref{lemma:2},
we can set the step sizes $\theta$ and $\lambda$ as  $\theta = \frac{\theta'}{2^m}$
and $\lambda = \frac{\lambda'}{2^k}$, and then
gradually increase the values of nonnegative integers $m$  and $k$
until relationships $f(x^{(T)}+\frac{\theta'}{2^m} y)> f(x^{(T)})$ and
$g(\alpha^{(T)}+\frac{\lambda'}{2^k} z)> g(\alpha^{(T)})$ are satisfied.
In this way, the following two theorems  can guarantee that
the   value of objective function $\mathcal{L}$ will
become  greater in the next iteration of Steps 2~\&~3.

\begin{theo} \label{theo:1}
Let  $x^{(T)}$  and $\alpha^{(T)}$ be the current values of $x$ and  $\alpha$, respectively,
and $m$ be a great enough nonnegative integer such that $f(x^{(T)}+\frac{\theta'}{2^m} y)< f(x^{(T)})$,
 if we update $x^{(T)}$ to $x^{(T+1)}$ by Eq.~(\ref{eq:step2}), then
we have
  \begin{align}
    \mathcal{L}(x^{(T+1)}, \alpha^{(T)}) \geqslant \mathcal{L}(x^{(T)}, \alpha^{(T)}),
  \end{align}
where the    equal sign holds if and only if $x^{(T)} = x^{(T+1)}$.
\end{theo}
 \textbf{Proof:} 
After updating,
 $x^{(T+1)} = x^{(T)}+\frac{\theta'}{2^m} y$.
If $y=0$, then $x^{(T+1)}=x^{(T)}$ and
%\begin{equation}
  $\mathcal{L}(x^{(T+1)}, \alpha^{(T)} ) = \mathcal{L}(x^{(T)},\alpha^{(T)})$;
%\end{equation}
% if $y\neq 0$, according to Lemma \ref{lemma:1}, we have
if $y\neq 0$, according to Lemma~1, we have
\begin{equation}
\begin{aligned}
   \mathcal{L}(x^{(T+1)}, \alpha^{(T)}) = & \mathcal{L}(x^{(T)}+\frac{\theta'}{2^m} y, \alpha^{(T)}) \\
  =& f(x^{(T)}+\frac{\theta'}{2^m} y)  \\
  > & f(x^{(T)}) \\
  =&\mathcal{L}(x^{(T)},\alpha^{(T)}).
\end{aligned}
\end{equation}
Therefore, Theorem 1 is correct.
\hfill $\blacksquare$

%After updating,
%  $x^{(T+1)} = x^{(T)}+\frac{\theta'}{2^m} y$.
% If $y=0$, then $x^{(T+1)}=x^{(T)}$ and
% %\begin{equation}
%   $\mathcal{L}(x^{(T+1)}, \alpha^{(T)} ) = \mathcal{L}(x^{(T)},\alpha^{(T)})$;
% %\end{equation}
% if $y\neq 0$, according to Lemma \ref{lemma:1}, we have
% \begin{equation}
% \begin{aligned}
%    \mathcal{L}(x^{(T+1)}, \alpha^{(T)}) = & \mathcal{L}(x^{(T)}+\frac{\theta'}{2^m} y, \alpha^{(T)}) \\
%   =& f(x^{(T)}+\frac{\theta'}{2^m} y)  \\
%   > & f(x^{(T)}) \\
%   =&\mathcal{L}(\alpha^{(T)},x^{(T)}).
% \end{aligned}
% \end{equation}
% Therefore, the theorem is correct.
% \hfill $\blacksquare$

\begin{theo} \label{theo:2}
Let  $x^{(T)}$  and $\alpha^{(T)}$ be the current values of $x$ and  $\alpha$, respectively,
 and
$k$ be a great enough nonnegative integer such that $g(\alpha^{(T)}+\frac{\lambda'}{2^k} z)> g(\alpha^{(T)})$,
 if we update $\alpha^{(T)}$ to $\alpha^{(T+1)}$ by Eq.~(\ref{eq:step3}), then
we have
  \begin{align}
    \mathcal{L}(x^{(T)}, \alpha^{(T+1)}) \geqslant \mathcal{L}(x^{(T)}, \alpha^{(T)}),
  \end{align}
where the    equal sign holds if and only if $\alpha^{(T)} = \alpha^{(T+1)}$.
\end{theo}
 \textbf{Proof:}
After updating,
 $\alpha^{(T+1)} = \alpha^{(T)}+\frac{\lambda'}{2^k} z$.
If $z=0$, then $\alpha^{(T+1)}=\alpha^{(T)}$ and
%\begin{equation}
  $\mathcal{L}(x^{(T)}, \alpha^{(T+1)} ) = \mathcal{L}(x^{(T)},\alpha^{(T)})$;
%\end{equation}
% if $z\neq 0$, according to Lemma \ref{lemma:2}, we have
if $z\neq 0$, according to Lemma~2, we have
\begin{equation}
\begin{aligned}
   \mathcal{L}(x^{(T)}, \alpha^{(T+1)}) = & \mathcal{L}(x^{(T)}, \alpha^{(T)}+\frac{\lambda'}{2^k} z) \\
  =& g(\alpha^{(T)}+\frac{\lambda'}{2^k} z)  \\
  > & g(\alpha^{(T)}) \\
  =&\mathcal{L}(x^{(T)},\alpha^{(T)}).
\end{aligned}
\end{equation}
Therefore, Theorem 2 is correct. \hfill $\blacksquare$

% The proof  is similar to  the proof   of Theorem \ref{theo:1}, while using Lemma \ref{lemma:2} instead of Lemma \ref{lemma:1}.
% \hfill $\blacksquare$

With the selected update directions and step sizes,
the problem in Eq.~(\ref{opt:gold}) can be solved by applying the
alternating maximization method.
%%%
Let $(x^*,\alpha^*)$ be the solution found by this method, then
$x^*$ indicates our estimated probabilities of the existence of
each potential parent-child influence relationship.
As the original problem in Eq.~(\ref{problem1})
aims at an integral solution for $x$,
we carry out the following two steps:
(1) By repeatedly sampling the value of  each $x_{ji}\in \{0,1\}$ in $x$
from probability $x^*_{ji}\in [0,1]$,
we obtain a set of samples $\{\hat{x}_1,...,\hat{x}_r\}$ for $x$, where $r$ is the times of sampling;
 (2) we select an optimal sample $\hat{x}^*$  from $\{\hat{x}_1,...,\hat{x}_r\}$ by solving
\begin{equation}
  \hat{x}^* = \arg \max \nolimits_{\hat{x}_i,i\in\{1,...,r\}} \mathcal{L}(\hat{x}_i, \alpha^*).
\end{equation}
Then,  $\hat{x}^*$ is our estimated integral solution for $x$.

\subsection{Pruning of Candidate Parent Nodes}
Given the fact that the infections of
nodes are only caused by their parent nodes with a certain
probability, the infections of the parent nodes and corresponding
child nodes should have relatively great positive
correlations. In contrast, if the infection statuses of two nodes
have no or an extremely low positive correlation, there is a
very low probability that these two nodes have an influence
relationship between them.
To quantify the  correlations of node infections, mutual information (abbreviated as
$MI$) is a commonly used criterion \cite{twind}. In our problem, it can be calculated as
\begin{equation} \label{eq:MI}
\begin{aligned}
  & MI(X_i,X_j) \\
  %= & \sum_{a=0}^1\sum_{b=0}^1 MI(X_i\!=\!a,X_j\!=\!b) \\
  = & \sum_{a=0}^1\sum_{b=0}^1 p(X_i\!=\!a,X_j\!=\!b)\ln \frac{p(X_i\!=\!a,X_j\!=\!b)}{p(X_i\!=\!a)p(X_j\!=\!b)},
\end{aligned}
\end{equation}
%where $MI(X_i=a,X_j=b)=p(X_i=a,X_j=b)\ln \frac{p(X_i=a,X_j=b)}{p(X_i=a)p(X_j=b)}$.
where $X_i$ is the infection status variable of node $v_i$,
\begin{equation}
  \begin{aligned}
    p(X_i\!=\!a)& = \frac{1}{\beta}\sum_{\ell=1}^{\beta} p(X_i^\ell=a),\\
    p(X_j\!=\!b)&=\frac{1}{\beta}\sum_{\ell=1}^{\beta} p(X_j^\ell=b),\\
    p(X_i\!=\!a,X_j\!=\!b)&=\frac{1}{\beta}\sum_{\ell=1}^{\beta} p(X_i^\ell=a)p(X_j^\ell=b),
  \end{aligned}
\end{equation}
and   $p(X_i^\ell\!=\!0)=1\!-\!s_i^\ell$,
    $p(X_i^\ell\!=\!1)= s_i^\ell$.

A greater $MI$ value  indicates a stronger correlation
between the infection statuses of nodes $v_i$ and $v_j$.
For a node   that has no influence relationship with $v_i$,
%%%%
its infection status often has no (or very low)
correlations to the infection status of $v_i$, resulting in a very
small $MI$ value  (close to 0).

Inspired by this line of reasoning, we screen out insignificant
candidate parent nodes for each node by the following two
steps. First, we calculate the $MI$ value for each two
nodes in the network. Then, we perform a modified
$K$-means algorithm with $K \!= \!2$ and one of the two means
fixed at 0 through all iterations of $K$-means, to efficiently
partition all MI values  into two groups,
where one group has a small mean close to 0.
Let $\eta$ be the
largest value in the group with a mean close to 0. Then, for
each $MI(X_i,X_j)  \leqslant \eta$, we regard the corresponding node $v_j$ as
an insignificant candidate parent node for node $v_i$ and exclude
 $v_j$ from the candidate parent node set $C_i$ of $v_i$.

 This heuristic pruning method should be performed before the estimation of influence relationship,
 to screen out
insignificant candidate parent nodes from the set $C_i$ of the candidate parent
nodes for each node $v_i$,
and  enables the
PIND algorithm to focus on influence relationships  that
are more likely to exist in the real diffusion network.

\subsection{Complexity Analysis}

 PIND algorithm
consists of the following two parts.
(1)
In the phase of  pruning   candidate parent nodes,
calculating $MI$ values requires $O(\beta n^2)$ time, and performing $K$-means on these $MI$ values takes $O(\tau n^2)$ time,  where $\beta$ is the number of historical diffusion processes, $n$ is the number of network nodes, and
$\tau$ is the number of iterations of $K$-means.
(2)
In the phase of estimating influence relationship,
the optimization problem in Eq.~(\ref{opt:gold}) is iteratively solved  by an alternating
maximization method. In each iteration of this method,  %the most computationally expensive process is
 calculating the partial derivatives can be finished within   $O(\beta c^2 n)$ time,
 and updating variables $x$ and $\alpha$ takes about $O(c n)$ time,
 %%%%%
 where  $c$ is the upper bound of the number of candidate parent nodes of each node in the network,
i.e., $c= \max\{|C_i|\mid i=1,...,n\} \ll n$. Therefore, solving the  problem in Eq.~(\ref{opt:gold}) takes $O(t \beta c^2 n)$ time, where $t$ is the number of iterations of   alternating
maximization  method.

In summary, the overall time complexity of PIND algorithm is $O(\beta n^2 + \tau n^2 + t \beta c^2 n)$.

\section{Experimental Evaluation}  \label{sec:experiment}
In this section, we first introduce the experimental setup, and
then evaluate our  PIND algorithm
  on synthetic and real-world networks. We investigate the effects of diffusion network size,
  the average degree of diffusion
network, the uncertainty of observed infection data, and the amount of
diffusion processes, on the
accuracy and running time of PIND. All algorithms
in the experiments are implemented in Python, running on a
MacBook Pro with Intel Core i5-1038NG7 CPU at 2.00GHz and 16GB RAM.
The source code of PIND and the data used in the experiments
are available at https://github.com/DiffusionNetworkInference/PIND.

\subsection{Experimental Setup}
\textbf{Network}.
We adopt LFR benchmark graphs \cite{lrf}  as  the synthetic  diffusion  networks.
 By setting different generation parameters, such as the number $n$ of nodes and the average degree of each node,
we generate a series of LFR benchmark graphs
with properties summarized in
%  in Table~\ref{uciDataSet}.
Table~\ref{uciDataSet}. % summarizes the properties of these  graphs.
 %%%
 Similar synthetic      network generation methods are commonly used  in existing studies \cite{twind,sidn,poind,tends}.
In addition, we also adopt  two commonly used  real-world microblogging networks \cite{11},  namely, (1) DUNF, which
contains 750 users and 2974 following relationships, and
(2) DPU, which contains 1038 users and 11385 following relationships.

\begin{table}[t]   \small
\caption{Properties of   LFR benchmark graphs  \label{uciDataSet}}
\begin{center}
  \begin{tabular}{|c|c|c|}
    \hline
    \textbf{Graphs} & \textbf{Number}  $\bm{n}$ \textbf{of Nodes} & \textbf{Average Degree} \\ \hline
    G1--G5 & 1000, 1500, 2000, 2500, 3000 & 4 \\ \hline
    G6--G10 & 2000 & 2, 3, 4, 5, 6 \\ \hline
  \end{tabular}
\end{center}
\end{table}

\textbf{Infection Data}.
The diffusion results $S=\{S_1,...,S_\beta\}$
can be generated by simulating $\beta$ times of diffusion processes on  each  network with randomly
selected  initially infected nodes in each simulation (the ratio of initially infected nodes
is 15\%). In each diffusion process, %($\ell \in\{1,...,\beta\}$),
 each infected node tries to
 infect its uninfected child nodes with a certain probability,
 which subjects to a Gaussian distribution with a mean of $0.3$ and
 a standard deviation  of 0.05, to make about 95\% of  infection propagation probabilities
within a range from 0.2 to 0.4.
Besides diffusion results $S$, the  cascades (i.e., the exact times when
node infections occurred) are also recorded for
cascade-based tested algorithms in the experiments.
Similar generation methods for infection   data
are commonly used in existing studies  \cite{1,06,11,twind,sidn,lmdn,tends,poind, grouplevel}.
%%Performance Criterion. To
%%
To add uncertainty into the   infection data,
for each exact node infection status $s \in\{0,1\}$, we replace it with $|s - u|$, where $u$ is a random uncertainty factor  and
its value subjects to a Gaussian distribution with a mean $\mu$  and
 a standard deviation  of 0.1 (if $\mu =0$, the standard deviation is  0).
All generated  infection  data are 
 stored  by OceanBase \cite{oceanbase2}.
%%%
 %Besides diffusion results $S$, the corresponding cascades (i.e., infection timestamps) are also recorded for
 %cascade-based tested algorithms in the experiments.

\textbf{Performance Criterion}.
We evaluate the performance of    {PIND}   algorithm in terms of the accuracy of structure inference.
For an inferred diffusion network,
     the accuracy  of structure inference   can be measured by
     the  F-score  of the  inferred directed edges:
     % \[ F\emph{-}score = \frac{2\cdot precision \cdot recall}{precision + recall},\]
     %       \[                     precision  =  \frac{N_{TP}}{N_{TP} + N_{FP}}, ~~ recall = \frac{N_{TP}}{N_{TP} + N_{FN}},\]
$F\emph{-}score = \frac{2\cdot precision \cdot recall}{precision + recall}$,
          where                     $precision  =  \frac{N_{TP}}{N_{TP} + N_{FP}}$, $recall = \frac{N_{TP}}{N_{TP} + N_{FN}}$,
      $N_{TP}$ is the number of true positives, i.e., the true edges which are correctly inferred by the algorithm;
     $N_{FP}$ is  the number of false positives, i.e., the wrong inferred edges which are not in the real network;
     and  $N_{FN}$ is the number of false negatives, i.e., the true edges which are not correctly inferred by the algorithm.
     On each network, we execute PIND 10 times,  and report the average F-score
     as the accuracy   of PIND on this network (the corresponding standard deviations are always less than 0.001 in all experiments).

     %A greater value of F-score indicates a more  accurate result of structure inference.

\textbf{Benchmark Algorithms}.
We compare   PIND   with (1) CORMIN \cite{12},  which is,  to  the best of our knowledge, the
 only existing approach to diffusion network inference with probabilistic data,
 (2) NetRate \cite{3},   which is a classical cascade-based
approach to   diffusion network inference,
 and    (3)
 TENDS \cite{tends}, which is a high-performance node infection status-based approach to  diffusion network inference.
%%%
In  our PIND algorithm, the number $r$ of  sampling rounds for $x$ is set to 100, %the maximum number $T$ of  iterations for  is set to 15,
and the stop condition for the iterative updates of $x$ and $\alpha$ is that the variations of each
$x_{ji}\in x$ and each $\alpha_{ji}\in \alpha$ are less than 0.01.
%%%
Since CORMIN requires to know
the node statuses at different timestamps,
we provide it with  the
temporal information of node infections.
%%%%
Since NetRate takes only cascades as input,
the uncertainties of node infection statuses do not work for NetRate. Therefore,
we execute NetRate with exact cascades in the experiments.
%%%%
Since  TENDS cannot directly deal with probabilistic data, we repeatedly sample 0/1 value 50 times from each probability $s^\ell_i$ ($i\in\{1,...,n\}$, $\ell\in \{1,...,\beta\}$)
to obtain 50 groups of samples of exact node infection statuses, %$\{\hat{S}_1,...,\hat{S}_{10}\}$,
and then report the best   F-score of TENDS on  these 50 groups of samples as the  accuracy of this approach.

 \subsection{Effect of Diffusion Network Size}

To study the effect of diffusion network size on algorithm performance, we
adopt     five synthetic  networks, i.e., G1--G5, whose sizes
 vary from 1000~to~3000.
We  simulate 300 times of diffusion processes on each network (i.e., $\beta = 300$), and set the  mean  $\mu$ of   uncertainty factor to 0.3.

\begin{figure}
	\centering
	\begin{subfigure}[b]{0.22000\textwidth}
		\centering
		\includegraphics[width=\textwidth]{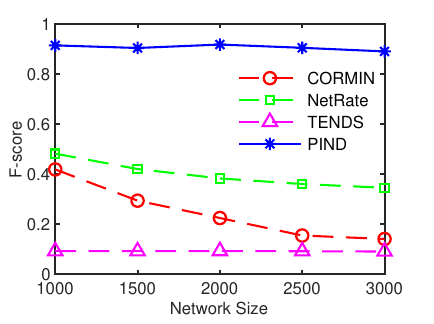}
		\caption{Accuracy}
		%\label{fig:tlarge_rlarge_dlarge_nhigh}
	\end{subfigure}
	\hfill
	\begin{subfigure}[b]{0.22000\textwidth}
		\centering
		\includegraphics[width=\textwidth]{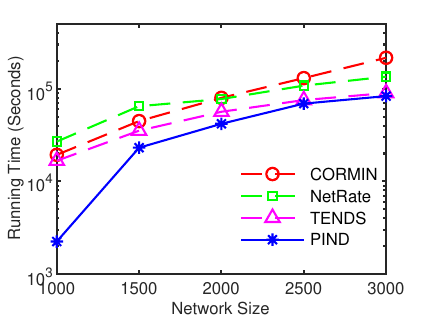}
		\caption{Running time}
		%\label{fig:tlarge_rlarge_dlarge_nlow}
	\end{subfigure}
	\caption{Effect of diffusion network size \label{fig:EffectOfSize} }
\end{figure}

Fig.~\ref{fig:EffectOfSize} illustrates   the F-score and execution time of each
 tested algorithm,
%%%
from  which we can observe  that PIND outperforms  CORMIN, NetRate and TENDS in terms of accuracy, and we can also have the following    observations.

(1) The accuracy of PIND and TENDS is reasonably insensitive to diffusion network size.
%%%
TENDS's insensitivity to diffusion network size has also been verified in existing study \cite{tends}.
Nonetheless, TENDS's accuracy in this experiment is poor. This is because
the samples of node infection statuses are far from the real situation, leading to large
inference errors for TENDS.
%%%%
This observation indicates that  straightforward sampling from the probabilistic data  is not a very
suitable approach to dealing with probabilistic data.
%%%
In contrast, in our PIND algorithm,  the theoretical guarantee on the improvement of its solution quality
helps it achieve a reasonably robust accuracy on networks with different sizes.

(2) Larger  network sizes   degrade the accuracy of CORMIN and NetRate.
This is because CORMIN and NetRate infer influence relationships by checking whether
the infections of two nodes are often within a time interval in historical diffusion processes.
Nonetheless, a few nodes may be infected at nearly the same time, even though they have no direct influence relationship.
In a larger   network,
 more nodes are likely to be involved in each diffusion process,
%the above situation  will appear  more often,  and   result  in more false positives in the inference results of CORMIN and NetRate.
 causing more aforementioned phenomena, which will  result in more false positives in the inference results of CORMIN and NetRate.

(3) The running time of each   algorithm increases with the  diffusion network size. PIND has better running time performance than CORMIN and
 NetRate.
 Although TENDS shows comparable efficiency to PIND on larger diffusion networks, it has a poor accuracy
as it has no effective strategy to  resist to
 the uncertainty in infection data.

 % is fastest as it has no aspect label estimation.

In addition, based on extensive testing on larger networks,
we have found that the running time of CORMIN and NetRate
rapidly increase with the growth of network size, and soon
exceed acceptable levels. Given this fact, we have selected
networks with sizes varying from 750 to 3000. Most existing
related work adopt similar network sizes.

\begin{figure}
	\centering
	\begin{subfigure}[b]{0.22000\textwidth}
		\centering
		\includegraphics[width=\textwidth]{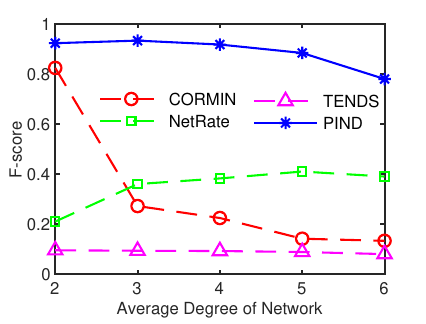}
		\caption{Accuracy}
		%\label{fig:tlarge_rlarge_dlarge_nhigh}
	\end{subfigure}
	\hfill
	\begin{subfigure}[b]{0.22000\textwidth}
		\centering
		\includegraphics[width=\textwidth]{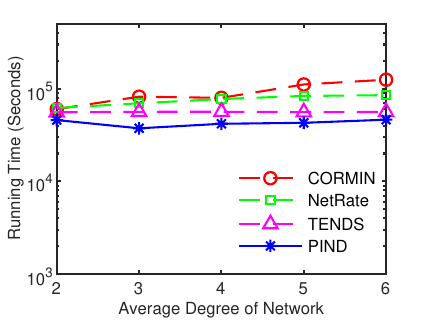}
		\caption{Running time}
		%\label{fig:tlarge_rlarge_dlarge_nlow}
	\end{subfigure}
	\caption{Effect of network's average degree     \label{fig:EffectOfDegree} }
\end{figure}

 \subsection{Effect of Network's Average Degree  }
 % The edge density of diffusion network can affect the number of influence relationships. The average node degree, i.e., the total number of edges divided by the total number of nodes, is usually used to represent the edge density of a network.
 
To study the effect of the average degree of diffusion network on algorithm performance, we
test the algorithms on  five synthetic  networks with the same size, i.e., G6--G10, whose average degree
varies from 2 to 6.
We  simulate 300 times of diffusion processes on each network (i.e., $\beta = 300$), and set the  mean  $\mu$ of    uncertainty factor to 0.3.

Fig.~\ref{fig:EffectOfDegree} illustrates the F-score and running time of each
tested algorithm, from which we can observe that PIND has the best accuracy compared with
   other tested algorithms,
and we can also have the following    observations.

(1)
With the increase of the average degree  of diffusion network,
the accuracy of CORMIN and PIND decrease.
The accuracy of NetRate
increases when the average degree increases from 2 to 5 and
then tends to decrease when the average degree reaches 6.
The reason behind is that
a greater average degree often brings more complicated influence
relationships between nodes, and thus adding complexity to the task of diffusion network inference.

(2) The running time of each
tested algorithm increases with the growth of average degree,
and PIND  is faster than CORMIN, NetRate and TENDS.

\subsection{Effect of Infection Data Uncertainty}

To study the effect of the uncertainty of infection data on algorithm  performance,
we test the algorithms on two real-world networks, i.e., DUNF and  DPU,
varying the  mean  $\mu$ of    uncertainty factor from 0 to 0.3 (with $\beta = 200$).

\begin{figure}
	\centering
	\begin{subfigure}[t]{0.22000\textwidth}
		\centering
		\includegraphics[width=\textwidth]{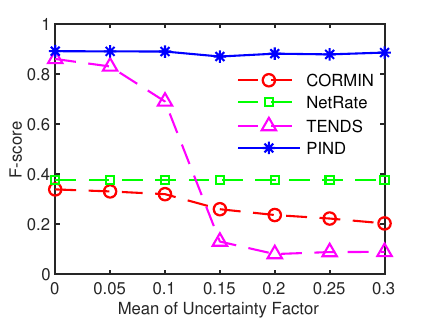}
		\caption{Accuracy on DUNF }
		%\label{fig:tlarge_rlarge_dlarge_nlow}
	\end{subfigure}
    \hfill
	\begin{subfigure}[t]{0.22000\textwidth}
		\centering
		\includegraphics[width=\textwidth]{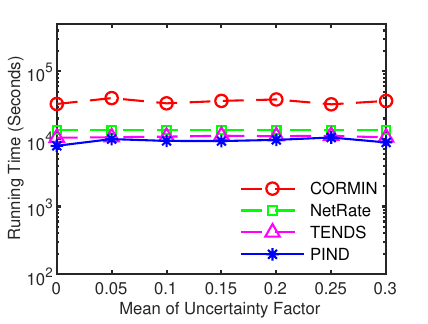}
		\caption{Running time on DUNF }
		%\label{fig:tlarge_rlarge_dlarge_nlow}
	\end{subfigure}
    \hfill
	\begin{subfigure}[t]{0.22000\textwidth}
		\centering
		\includegraphics[width=\textwidth]{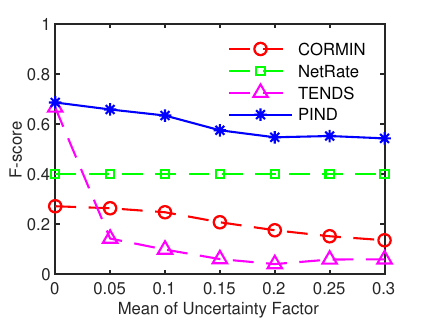}
		\caption{Accuracy  on DPU}
		%\label{fig:tlarge_rlarge_dlarge_nhigh}
	\end{subfigure}
	\hfill
	\begin{subfigure}[t]{0.22000\textwidth}
		\centering
		\includegraphics[width=\textwidth]{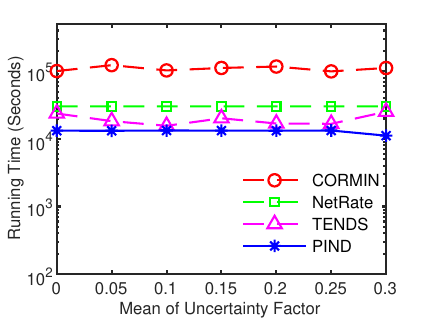}
		\caption{Running time on DPU}
		%\label{fig:tlarge_rlarge_dlarge_nlow}
	\end{subfigure}
	\caption{Effect of infection data  uncertainty   \label{fig:EffectOfInitial222} }
\end{figure}

\iffalse
\begin{figure}
	\centering
	\begin{subfigure}[t]{0.22000\textwidth}
		\centering
		\includegraphics[width=\textwidth]{figures/uncertainty_f1_dpu.pdf}
		\caption{Accuracy  }
		%\label{fig:tlarge_rlarge_dlarge_nhigh}
	\end{subfigure}
	\hfill
	\begin{subfigure}[t]{0.22000\textwidth}
		\centering
		\includegraphics[width=\textwidth]{figures/uncertainty_time_dpu.pdf}
		\caption{Running time }
		%\label{fig:tlarge_rlarge_dlarge_nlow}
	\end{subfigure}
	\caption{Effect of uncertainty of observation data  on DPU \label{fig:EffectOfInitial111} }
 \end{figure}
\fi

Fig.~\ref{fig:EffectOfInitial222} %~\&~\ref{fig:EffectOfInitial111}
 illustrates the F-score and running time of each
tested algorithm on DUNF and DPU.
From the figure we  can have the following observations.

(1)  Compared with   other tested algorithms,
PIND often shows a significant advantage on accuracy,
while TENDS achieves a very close accuracy to PIND  when  $\mu \leqslant 0.05$ on DUNF  and when $\mu = 0 $ on DPU.
The reason behind is  that
with a small enough mean of uncertainty factor,
the sampled node infection statuses for TENDS will be equal or close enough to the real
situation. With correct infection data,   TENDS is able to
achieve a high accuracy performance.
%The effectiveness of TENDS with  correct node infection statuses
%has been verified  in existing study \cite{sidn}.

(2) A higher uncertainty tends to degrade the accuracy of PIND, CORMIN and TENDS.
This is because
a higher uncertainty  makes measuring
the correlations between node infections     more difficult.
When the uncertainty exceed a threshold, for example, $\mu\geqslant 0.5$,
the distinction between infected and uninfected statues will
totally lose, then it becomes almost
impossible to recover the real influence relationships.
%%On the other hand, a higher uncertainty does not show
Note that the accuracy of NetRate does not change with uncertainty, since NetRate
uses exact cascades  for diffusion network inference in this experiment.

(3) PIND outperforms CORMIN in running time performance,
and is often relatively faster than NetRate and TENDS. In addition,
the uncertainty has mild effect on the running time of tested algorithms.
This is because the time complexity of each tested algorithm is mainly dominated by
the network size and the amount of diffusion processes.

%Similar results can   be observed on synthetic networks.

\subsection{Effect of   Diffusion Process  Amount }
% The inference of a diffusion network is based on the observed results of diffusion processes. Hence, the amount of diffusion processes may affect the accuracy of diffusion network inference. Generally, more diffusion processes will expose more information about a diffusion network, and this may help diffusion network inference algorithms achieve more accurate inference results.

%More diffusion processes   provide more infection data for diffusion network inference.
To  study the effect of diffusion process amount on algorithm  performance,
we test the algorithms on DUNF and  DPU with different number $\beta$ of diffusion processes, varying
from 100 to 300. For the diffusion results obtained
with each $\beta$, we set   the mean
$\mu$ of uncertainty factor to 0.3.

\begin{figure}
	\centering
	\begin{subfigure}[t]{0.22000\textwidth}
		\centering
		\includegraphics[width=\textwidth]{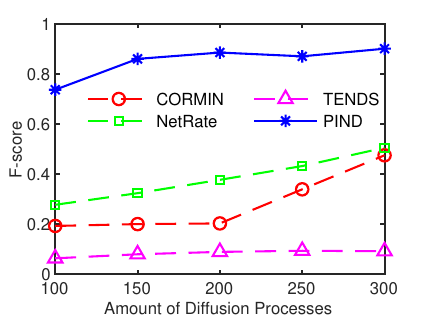}
		\caption{Accuracy on DUNF }
		%\label{fig:tlarge_rlarge_dlarge_nlow}
	\end{subfigure}
    \hfill
	\begin{subfigure}[t]{0.22000\textwidth}
		\centering
		\includegraphics[width=\textwidth]{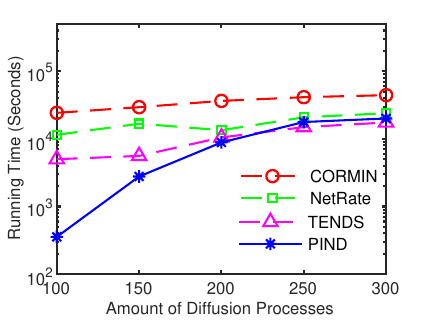}
		\caption{Running time on DUNF }
		%\label{fig:tlarge_rlarge_dlarge_nlow}
	\end{subfigure}
\hfill
    	\begin{subfigure}[t]{0.22000\textwidth}
		\centering
		\includegraphics[width=\textwidth]{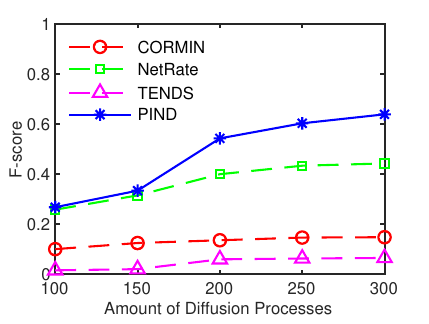}
		\caption{Accuracy on DPU }
		%\label{fig:tlarge_rlarge_dlarge_nhigh}
	\end{subfigure}
	\hfill
	\begin{subfigure}[t]{0.22000\textwidth}
		\centering
		\includegraphics[width=\textwidth]{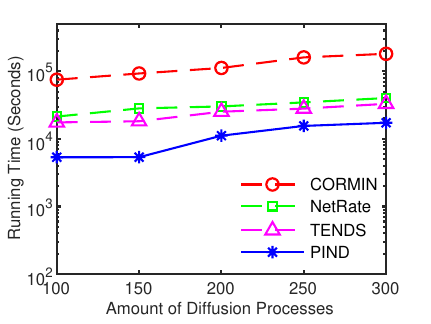}
		\caption{Running time on DPU  }
		%\label{fig:tlarge_rlarge_dlarge_nlow}
	\end{subfigure}
	\caption{Effect of  the  amount of diffusion processes \label{fig:EffectOfBeta222} }
\end{figure}

\iffalse
\begin{figure}
	\centering
	\begin{subfigure}[t]{0.22000\textwidth}
		\centering
		\includegraphics[width=\textwidth]{figures/amount_f1_dpu.pdf}
		\caption{Accuracy  }
		%\label{fig:tlarge_rlarge_dlarge_nhigh}
	\end{subfigure}
	\hfill
	\begin{subfigure}[t]{0.22000\textwidth}
		\centering
		\includegraphics[width=\textwidth]{figures/amount_time_dpu.pdf}
		\caption{Running time  }
		%\label{fig:tlarge_rlarge_dlarge_nlow}
	\end{subfigure}
	\caption{Effect of the   amount of diffusion processes  on DPU \label{fig:EffectOfBeta111} }
\end{figure}
\fi

Fig.~\ref{fig:EffectOfBeta222} %~\&~\ref{fig:EffectOfBeta111}
illustrates the F-score and running time of
each tested algorithm on DUNF and DPU. From
the figure we can %observe that PIND achieves the best accuracy compared with the other algorithms, and
%we can also
 have the following    observations.

(1) A greater amount of  diffusion processes often
helps the   algorithms achieve higher accuracy.
This is because
 more diffusion processes tend to involve more nodes, and  activate more influence relationships
 (i.e., infections spread through these influence relationships), which enable the algorithms  to
 learn a  more complete  network structure.

(2) Compared with   other algorithms, PIND achieves significantly  better accuracy
in all  but two settings, i.e., $\beta =100$ and $\beta =150$ on DPU.
%%%%%%
This is because
when $\beta \leqslant 150$,
the amount of diffusion processes  are insufficient
to activate enough influence relationships in DPU, and thus degrade the accuracy of PIND  in these settings.
%In this kind of situation,
%the exact cascades used by NetRate helps the approach achieve

(3)  The running time of each algorithm increase with the amount of diffusion processes.
PIND has significantly
better running time performance than CORMIN,
and tends to be relatively more efficient than  NetRate.
 Although TENDS
shows comparable (or even slightly better) efficiency to PIND on larger diffusion
networks, it suffers from a poor accuracy as it can not effectively deal with probabilistic data.

%Similar results can   be observed on synthetic networks.

\subsection{Effect of Iterations of PIND }

To study the effect of  iterations on the performance of PIND,
we execute PIND on DUNF and DPU (with  $\beta=200$, $\mu = 0.3$) and
report the accuracy of its latest inference result   at the end of each iteration.

\begin{figure}
	\centering
	\begin{subfigure}[t]{0.22000\textwidth}
		\centering
		\includegraphics[width=\textwidth]{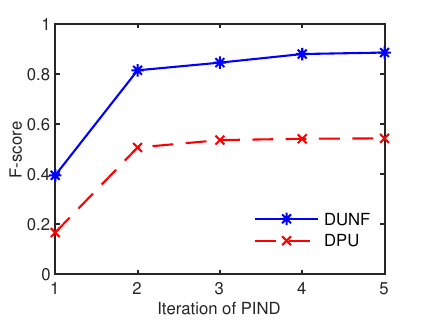}
		\caption{Accuracy  }
		%\label{fig:tlarge_rlarge_dlarge_nhigh}
	\end{subfigure}
	\hfill
	\begin{subfigure}[t]{0.22000\textwidth}
		\centering
		\includegraphics[width=\textwidth]{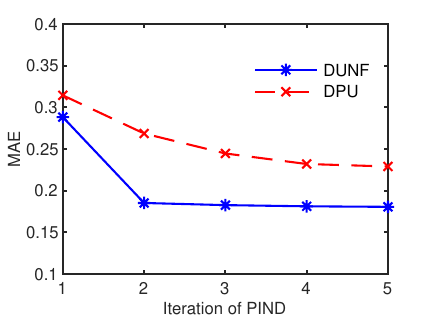}
		\caption{MAE of learned $\bm \alpha_{ji}$  }
		%\label{fig:tlarge_rlarge_dlarge_nlow}
	\end{subfigure}
	\caption{Effect of the  iterations of PIND \label{fig:EffectOfIter} }
\end{figure}

 Fig.~\ref{fig:EffectOfIter}~(a) illustrates the F-score of PIND at the end of each iteration,
 from which  we can observe that
  the  accuracy of PIND can be  improved with more iterations,
 and  shows a fast convergence property.

 The direct reason behind this  is
 that PIND can accurately estimate  the
existence of influence relationships. %(i.e., $\{x_{ji} \mid v_i \in V, v_j\in C_i\}$).
Another
important reason is that PIND can accurately
infer infection propagation probabilities. %(i.e., $\{\alpha_{ji} \mid v_i \in V, v_j\in C_i\}$).
To demonstrate this
effectiveness, Fig.~\ref{fig:EffectOfIter}~(b) illustrates the MAE (Mean Absolute
Error) of infection propagation probabilities learned by PIND  at the end of each iteration.
A lower MAE indicates a higher accuracy.
From Fig.~\ref{fig:EffectOfIter}~(b), we can observe that
the MAE of learned $\alpha_{ji}$ can also be  improved with more iterations.
%%%%
For comparison,
the corresponding MAE values of  NetRate on
DUNF and DPU are 0.2822 and 0.2791, respectively, which are significantly higher than
that of PIND.

%==============================================================================
\section{Conclusion} \label{sec:conclusion}
In this paper,
we have investigated the problem of how to infer a
diffusion network using only the probabilistic data
of  the node  infection statuses observed in historical diffusion processes.
Towards this, we have formulated the problem as a constrained nonlinear regression w.r.t.~the probabilistic data,
and proposed an effective and efficient algorithm, PIND, to solve the regression problem
in an iterative way.
The improvement of solution quality in each iteration can be
theoretically guaranteed.
%%%
Extensive experimental results  on  both   synthetic and real-world networks have %been conducted, and the results have
verified  the effectiveness and efficiency of PIND. 

% So far we have applied our PIND  algorithm to infer influence relationships in diffusion networks that have static structures.
% For the next stage of study, a promising direction is to extend  our approach to handle with dynamic diffusion networks.

\section{ Acknowledgments}

This work was supported by the National Key Research and Development Program of China (2022YFB3105000), the National Natural Science Foundation of China (61976163), the Key R\&D Program of Hubei Province (2023BAB077), and the Fundamental Research Fund for   Central Universities (2042023kf0219). This work was supported by Ant Group.
Qian Yan is  the corresponding author.

%\bibliography{aaai24}

\end{document}